\documentclass[%
 aps,
 pre,%
 amsmath,amssymb,
reprint,%
author-numerical,%
]{revtex4-1}

\usepackage{graphicx}
\usepackage{dcolumn}
\usepackage{bm}
\usepackage[mathlines]{lineno}
\usepackage{amsmath}
\usepackage{url}

\usepackage{booktabs}
\usepackage{rotating}
\usepackage{multirow}
\usepackage{subfigure}
\usepackage[normalem]{ulem}
\usepackage{amsmath}
\usepackage[usenames,dvipsnames,svgnames,table]{xcolor}
\usepackage{booktabs}
\usepackage{rotating}
\usepackage{multirow}
\usepackage{subfigure}

\usepackage{color}
\usepackage{hyperref}
\hypersetup{
     colorlinks   = true,
     citecolor    = blue,
     linkcolor    = blue,
     urlcolor     = blue
}

\begin{document}

\title{General form of the tunneling barrier for nanometrically sharp electron emitters}

\author{Andreas Kyritsakis}
\email{akyritsos1@gmail.com; andreas.kyritsakis@ut.ee}
\affiliation{Institute of Technology, University of Tartu,  Nooruse 1, 50411 Tartu, Estonia}

\date{\today}

\begin{abstract}
 Field electron emission from nanometer-scale objects deviates from the predictions of the classical emission theory as both the electrostatic potential curves within the tunneling region and the image potential deviates from the planar one. This impels the inclusion of additional correction terms in the potential barrier. At the apex of a tip-like rotationally symmetric surface, these terms are proportional to the (single) local emitter curvature. The present paper generalizes this relation, showing that for any emitter geometry, the coefficient of the correction terms is given by the mean curvature, i.e. the average of the two principal curvatures. 
\end{abstract}

\maketitle

\section{Introduction}
The most fundamental step in deriving theories of thermal-field electron emission is writing an expression for the tunneling barrier potential energy $U(z)$, where $z$ is the distance measured normally from the emitter's electrical surface.
The first theory of field emission by Fowler and Nordheim \cite{FN1928} was based on the exact triangular barrier, while later theories \cite{Nordheim1928} included the image potential corrections to the barrier. 
The standard theory typically used nowadays, i.e. the one by Murphy and Good \cite{MurphyG}, as well its recent generalizations for the thermal-field regime \cite{Jensen2006}, use the standard image-corrected triangular barrier form (also known as the Schottky-Nordheim barrier), which for an electron with impinging energy at the Fermi level is
\begin{equation}
    \label{planarBarrier}
    U(z) = \phi - \frac{Q}{z} - e F z \textrm{.}
\end{equation}
In eq. \eqref{planarBarrier}, the zero-field barrier height is equal to the local work function $\phi$,  $F$ is the magnitude of the local  electrostatic field at the surface, $Q = e^2 / (16 \pi \epsilon_0) \approx 0.36 \textrm{ eV nm}$ is a universal constant, and $e$ is the elementary charge.

To derive this formula, the electrostatic potential has been approximated to be linear with the distance from the surface $z$, i.e. $\Phi(z) = Fz$, while the exchange and correlation interactions have been approximated by the planar image interaction.
Both of these approximations are consistent with a quasi-planar emitting surface, which is a valid approximation for emitters with radii of curvature larger than about 20 nm \cite{KXnonfn}.

However, it is well-known that this approximation is not valid for emitters with nm-scale radii of curvature, as has been shown both theoretically and experimentally \cite{CutlerAPL,KXnonfn,KXscaling,CabreraPRB,EdgcombeDeduction}.
In such cases, the electrostatic potential becomes curved within the tunneling region and the image potential slightly deviates from the planar one. This renders eq.~\eqref{planarBarrier} insufficient and the inclusion of correction terms necessary.

Kyritsakis and Xanthakis \cite{KXnonfn} used a quadratic expansion of the electrostatic potential and the spherical image interaction, yielding the following barrier form 
\begin{equation} \label{potentialQuad}
    U(z) = \phi - \frac{Q}{z+\frac{z^2}{2R}} - e F z + e F \frac{z^2}{R}  \textrm{.}
\end{equation}
We then derived a generalized Fowler-Nordheim-type equation for the local emission current density $J$ as a function $F, \phi$, and the emitter radius of curvature:
\begin{equation} \label{current}
\begin{aligned}
   J(F,\phi,R) = & a \frac{F^2}{\phi} \left[ \frac{1}{\lambda_d(f)} + \frac{\phi}{eFR} \psi(f) \right]^{-2} \\ 
   & \exp\left[b \frac{\phi^{3/2}}{F} \left(\nu(f) + \omega(f)\frac{\phi}{eFR} \right) \right] \textrm{.}
\end{aligned}
\end{equation}
In the above equation, $f \equiv (e^3/4 \pi \epsilon_0) (F/ {\phi}^2) = (1.439964 {\textrm{ eV}}^2 {\textrm{ V}}^{-1} \textrm{ nm) } (F/ {\phi}^2) $ is the reduced field strength, $\nu(f),\omega(f),\lambda_d(f), \psi(f)$ are known and tabulated functions \cite{KXnonfn}, and $a \equiv e^3 / (16 \pi^2 \hbar) \approx  1.541434 \times 10^{-6} \textrm{A eV V}^{-2}$, $b \equiv 4\sqrt{2 m}/3e\hbar \approx 6.830890 \textrm{ (eV)}^{-3/2} \textrm{ V nm}^{-1}$ are universal constants, also known as the first and second Fowler-Nordheim constants respectively.
The resulting current density vs field plot deviates significantly from the typical straight-line Fowler-Nordheim behavior, with a curvature that scales with the emitter curvature $1/R$.

It was also shown that at the apex of a typical rotationally symmetric emitting nanotip, which is an umbilic point \cite{Umbilic} of the emitting surface (has a single radius of curvature), the quadratic term of the electrostatic potential is inversely proportional to the (single) local radius of curvature, as in eq. \eqref{potentialQuad}.
Since in such tips, most of the emission is coming from the vicinity of the apex, approximating the quadratic term as the apex curvature yields a reasonably good approximation for the emission current.

However, non-tip-like emitters that are not rotationally symmetric and cannot be described by the above approximation have started being studied a lot. Especially edge-type emitters from two-dimensional materials such as nanosheets and nanoflakes have recently attracted significant interest \cite{santandrea2011field,giubileo_field_2019,di_bartolomeo_leakage_2016,giubileo_field_2023,iemmo_nanotip_2020,pelella_gate-controlled_2021,patra_field_2021,di_bartolomeo_leakage_2016}. 
Furthermore, modern numerical models of electron emission \cite{Eimre2015,GETELECpaper, kyritsakis2018thermal, veske2019dynamic} need to resolve the emission distribution at each point of the emitter separately. 
This becomes even more relevant for thermal-field and photo-excited Schottky-type electron sources \cite{reynolds2023nanosecond}, for which a significant proportion of the emission comes from off-axis regions of the emitting surface. These regions are also typically non-umbilic, since the azimuthal curvature deviates from the polar one as we move away from the apex.

In Refs. \cite{GETELECpaper,kyritsakis2018thermal}, this problem was tackled by calculating numerically the whole electrostatic potential function along the emission path. 
However, this is computationally expensive as it requires very high numerical accuracy in the tunneling region. 
Furthermore, it is not yet clear whether the relation between the non-planar correction for the image potential and the quadratic term of the electrostatic potential should be the same as in eq. \eqref{potentialQuad}.
Hence, deriving a general expression for the barrier that is valid for any emitting surface point (not only umbilic ones) is necessary for both the accurate theoretical calculation and the precise and computationally efficient simulation of the emitted current density from emission surfaces with arbitrary geometry.

In this paper, I derive a general asymptotic expansion for both the electrostatic potential and the image interaction, which are valid at any point of any continuous emitter surface. 
These results render eq. \eqref{current} valid for any emitter shape and show that the appropriate value for the quadratic parameter $R$ is the radius of mean curvature \cite{Curvature}, i.e. $R=R_m$. 
This result contradicts the previous findings of Biswas et. al. \cite{biswas_tunneling_2018}, who found that the appropriate value of the $R$-parameter for the ellipsoid and hyperboloid shapes is the second (smaller) principal radius of curvature. 
A brief revision of their derivation pinpoints a subtle mathematical error that yielded this mistake. 

\section{General description of the emitting surface}
\begin{figure}
    \centering
    \includegraphics{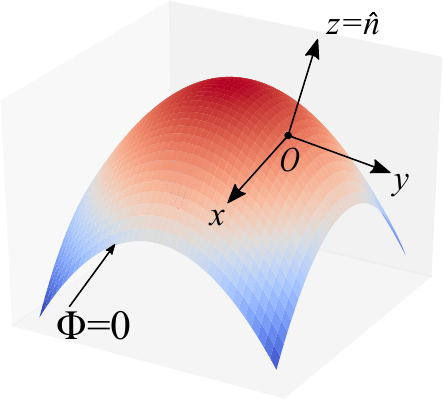}
    \caption{Schematic of the considered equipotential surface and coordinate system.}
    \label{schematic}
\end{figure}

Consider a generic emitting surface and an arbitrary point $O$ on it, as shown in Fig. \ref{schematic}. Without loss of generality, I define a Cartesian coordinate system centered at $O$, with the $z-$axis being perpendicular to the surface, i.e. $\hat{z} \equiv \hat{n}$, and $x,y$ being aligned with the principal curvature axes of the surface with $x$ being the one that corresponds to the higher principal curvature (smaller radius). 
This coordinate system is known in differential geometry as the Darboux frame of an arbitrary curve belonging to the surface. 
In the vicinity of $O$, the surface can be described by the Monge patch \cite{Monge} $\mathbf{r} = (x, y, -g(x,y))$, where $g(x,y)$ is a smooth function of $x,y$ and the minus sign is chosen to facilitate the convention that the curvature is considered positive if it is directed downwards. 
The perpendicular vector at a given point of the surface is given by 
\begin{equation}
    \hat{n} = \frac{\mathbf{r}_x \times \mathbf{r}_y}{|\mathbf{r}_x \times \mathbf{r}_y |} = \frac{(g_x,g_y,1)}{\sqrt{g_x^2+g_y^2+1}},
\end{equation}
where subscripts denote partial derivatives.
Given the selection of the coordinate system, $\hat{n} \equiv \hat{z}$, it is $g_x=g_y=0$ at $O$.
In the Monge patch representation of the surface, the mean curvature is given by \cite{Monge}
\begin{equation}
    H = - \frac{(1+g_y^2)g_{xx} - 2 g_x g_y g_{xy} + (1 + g_x^2)g_{yy}}{2(1+g_x^2+g_y^2)^{3/2}} \textrm{.}
\end{equation}
In the chosen coordinate system, the above expression evaluated at $O$, where $g_x=g_y=0$, simplifies into
\begin{equation}
    H(O) \equiv \frac{1}{R_m} = - \frac{g_{xx} + g_{yy}}{2} \textrm{,}
\end{equation}
where $R_m \equiv 1/H$ is the local radius of mean curvature.

\section{Electrostatic potential}

The most important part of the curvature-related corrections to the tunneling barrier comes from the electrostatic potential $\Phi(z)$.
In the barrier formula \eqref{potentialQuad}, the electrostatic potential is approximated as
\begin{equation}
    \label{electrostatic}
    \Phi(z) = Fz \left(1- \frac{z}{R} + \mathcal{O}\left( \frac{z}{R} \right)^2 \right) \textrm{, } z \ll R \textrm{.}
\end{equation}
The above formula is based on expanding the electrostatic potential in a Taylor polynomial 
\begin{equation}
    \Phi(z) = \Phi(0) + Fz + \frac{1}{2}\Phi_{zz}(O)z^2 + \mathcal{O} \left( z^3 \right) \textrm{,   } z \rightarrow 0
\end{equation}
and keeping up to the quadratic order term.
It is evident then that the curvature parameter $R$ is given by 
\begin{equation} \label{Radius}
    R = -\frac{2F}{\Phi_{zz}(O)} \textrm{,}
\end{equation}
where subscripts of functions denote the corresponding partial derivatives.
In the following section, I will show that in the general case of an arbitrary surface, $R=R_m$, i.e., it is equal to the local radius of the average curvature of the surface.

\subsection{Universal expansion}

Consider the potential along the $z$ axis $\Phi(x=0,y=0,z)$, assuming that the surface $(x,y,-g(x,y))$ is equipotential and ---without loss of generality-- grounded, i.e.
\begin{equation} \label{surface}
    \Phi(x,y,-g(x,y)) = 0 \textrm{.}
\end{equation}

Taking the derivatives of eq. \eqref{surface} with respect to $x$ yields
\begin{equation} \label{first_der1}
    \Phi_x - \Phi_z g_x = 0 \textrm{.}
\end{equation}
Writing the same equation for the $y-$derivatives and evaluating at $O$ where $g_x = g_x = 0$ yields
\begin{equation}
    \label{first_ders}
    \Phi_x(O) = \Phi_y(O) = 0 \textrm{.}
\end{equation}
In order to write the second $x-$derivatives of equation \eqref{surface}, eq. \eqref{first_der1} needs to be differentiated, giving
\begin{equation}
    \label{second_der1}
    \Phi_{xx} - 2g_x\Phi_{xz} + \Phi_{zz}g_x^2 - \Phi_z g_{xx} = 0 \textrm{.}
\end{equation}
Evaluating eq. \eqref{second_der1} at $O$, and performing the same calculations for the $y-$derivatives yields 
\begin{equation} \label{second_ders}
    \Phi_{xx}(O) = F g_{xx}(O) , \textrm{ }  \Phi_{yy}(O) = F g_{yy}(O) \textrm{.}
\end{equation}

Considering that the electron emission is occurring in a vacuum (disregarding any space charge effects), the electrostatic potential $\Phi$ satisfies the Laplace equation, i.e., 
\begin{equation}
    \label{laplace}
    \Phi_{zz} = - \Phi_{xx} - \Phi_{yy} \textrm{.}
\end{equation}
Substituting $\Phi_{xx}, \Phi_{yy}$ from \eqref{second_ders} yields
\begin{equation} \label{eq:central1}
    \Phi_{zz}(O) = -F(g_{xx} + g_{yy}) = -\frac{2F}{R_m} \textrm{,}
\end{equation}
which in view of \eqref{Radius} gives the central result of this section:
\begin{equation}
    \label{central}
    R = \frac{-2}{g_{xx}(O) + g_{yy}(O)} = \frac{1}{H(O)} = R_m
\end{equation}
Note that the above equation is general. The selection of the point $O$ is absolutely arbitrary and the only assumption about the shape of the equipotential surface is that it is mathematically smooth (twice differentiable).

\subsection{Error in the literature}
A comment is warranted on the result obtained for ellipsoid and hyperboloid emitters by Biswas et. al.~\cite{biswas_tunneling_2018}, which is contradicting the above general expression. Revisiting the derivation of reference~\cite{biswas_tunneling_2018}, it is evident that substituting their eq. (8) into the expression of the potential as a function of the spheroidal coordinates does not yield their eq. (10). 
The truncated $\mathcal{O}\left(\Delta s^2 \right)$ terms of their eq.~(8) should yield an $\mathcal{O} \left( \Delta s^2 \right)$ contribution, which has been completely disregarded. Considering this contribution properly would lead to the general result of eq.~\eqref{central}.

To confirm the latter and validate the main result of this paper, I shall calculate $R$ for the specific hyperboloid tip geometry, which is addressed in section IIB of Ref.~\cite{biswas_tunneling_2018}.
The electrostatic potential is given as a function of the prolate spheroidal coordinates $\eta, \xi$ (defined as in Ref.~\cite{biswas_tunneling_2018}) as 
\begin{equation} \label{eq:potential_hyperboloid}
    \Phi(\eta, \xi) = V \left( 1 - \frac{\log\left( \frac{1-\xi}{1+\xi} \right)}{\log\left( \frac{1-\xi_0}{1+\xi_0}\right) } \right) \textrm{,}
\end{equation}
where $\xi = \xi_0 > 0$ defines the equipotential surface $\Phi = 0$ of the emitter and $\xi=1$ defines the anode where $\Phi = V$.
The corresponding electric field perpendicular to the emitter surface is
\begin{equation} \label{eq:field_hyperboloid}
    F(\eta, \xi) = \frac{2V}{c\log\left( \frac{1-\xi}{1+\xi_0}\right)} \frac{1}{\sqrt{(1-\xi_0^2)(\eta^2 - \xi_0^2)}} \textrm{,}
\end{equation}
where $c$ is the focal length of the hyperboloid. 

As in the general case, I define the $z$-coordinate at an arbitrary point $(\eta, \xi_0)$ on the emitter surface as the distance from the point along the perpendicular line. 
Using the general definition of eq.~\eqref{Radius}, it yields
\begin{equation} \label{eq:kappa_hyperboloid}
    \frac{1}{R} = -\frac{1}{2F} \frac{\partial F}{\partial z} = -\frac{1 }{2F h_\xi} \frac{\partial F}{\partial \xi} \textrm{,}
\end{equation}
where $h_\xi = c \sqrt{\frac{\eta^2 - \xi^2}{1 - \xi^2}}$ is the metric factor. 
Evaluating eq.~\eqref{eq:kappa_hyperboloid} by differentiating~\eqref{eq:field_hyperboloid} yields
\begin{equation}
    R  = \frac{2c\left(\eta^2 - \xi_0^2\right)^{3/2}\sqrt{1-\xi_0^2}}{\xi_0 \left(1 - 2\xi_0^2 + \eta^2\right)}
\end{equation}

The principal radii of curvature of the emitter hyperboloid are~\cite{biswas_tunneling_2018}
\begin{equation}
    R_1 = \frac{c}{\xi_0} \frac{\left( \eta^2 - \xi_0^2 \right)^{3/2}}{\sqrt{1 - \xi_0^2}}
\end{equation}
\begin{equation}
    R_2 = \frac{c}{\xi_0} \sqrt{\left( \eta^2 - \xi_0^2 \right) \left(1 - \xi_0^2 \right)} \textrm{.}
\end{equation}
After a few algebraic manipulations, it yields
\begin{equation}
    R_m = \frac{1}{2} \left( \frac{1}{R_1} + \frac{1}{R_2} \right)  = \frac{2c\left(\eta^2 - \xi_0^2\right)^{3/2}\sqrt{1-\xi_0^2}}{\xi_0 \left(1 - 2\xi_0^2 + \eta^2\right)}
\end{equation}
which confirms the main result of this paper, i.e., $R = R_m$.

\section{The image interaction}
Apart from the electrostatic field, when an emitter becomes highly curved, the image potential is also modified compared to the planar one. 
In eq.~\eqref{potentialQuad}, the image potential
\begin{equation}
    \label{imagesphere}
    U_i(z) = -\frac{Q}{z+\frac{z^2}{2R}}
\end{equation}
has been approximated by that of a grounded sphere near a point charge, which is a common practice in modern field emission theories~\cite{EdgcombeDeduction, KXnonfn, KXscaling, edgcombe_currentvoltage_2003, Holgate2017, KXAPL,KXBeamspot, KXGTF}.
Within the derivation of the current density expressions of Ref.~\cite{KXnonfn}, which are based on asymptotic expansions for $z \ll R$, $U_i(z)$ can be approximated as 
\begin{equation}
    \label{imageSphereAsymptotic}
    U_i(z) = - \frac{Q}{z} \left( 1 - \frac{z}{2R} + \mathcal{O} \left( \frac{z}{R}\right)^2 \right) \textrm{, } z \ll R
\end{equation}
without any change in the final result.
In the following paragraphs, I will show that eq.~\eqref{imageSphereAsymptotic} is valid for any surface shape, with the appropriate parameter for $R$ being the radius of the mean curvature, i.e. $R = R_m$, similarly to the case of the electrostatic potential. 
The combination of this result with the one of eq.~\eqref{central} shows that the generalized emission equations derived by Kyritsakis and Xanthakis~\cite{KXnonfn, KXGTF} and used in modern computational models \cite{GETELECpaper} hold for any emitter surface geometry, as long as the parameter $R$ is substituted by the radius of the mean curvature $R_m$.

In order to prove this, consider the generic surface of Fig.~\ref{schematic} and a point charge $q$ residing on at the point  $\mathbf{r}_q=(0,0,z_q)$. 
In order to obtain the image interaction, we need to first solve the Poisson equation 
\begin{equation}
    \label{eq:Poisson}
     \nabla^2 \Phi  = - \frac{q}{\epsilon_0} \delta(x) \delta(y) \delta(z-z_q)
\end{equation}
with the boundary condition $\Phi = 0$ on the surface, where $\delta(\cdot)$ denotes the Dirac functional.
To find the image interaction energy, $\Phi$ is separated into the contribution of the point charge $\Phi_q = q/4 \pi \epsilon_0 |\mathbf{r} - \mathbf{r}_q|$,
and the contribution of the image charges $\Phi_{i} = \Phi - \Phi_q$. 
Then the potential energy of the interaction between the point charge and the image charges can be found by integrating the energy spent to introduce an infinitesimal charge $dq'$ at $\mathbf{r}_q$~\cite{jackson1975classical}
\begin{equation}
    \label{eq:ImageInteraction}
    U_i(q) = \int_0^q \Phi_i (\mathbf{r}_q,q')  dq' \textrm{.}
\end{equation}

We are now interested in the image interaction for $z_q \ll R_m$.
As the charge approaches the surface, the image charges on the surface accumulate around $O$ and the surface can be approximated by a flat plane.
In the following, I shall exploit this and consider the surface as a small perturbation from a plane in the vicinity of $O$, utilizing boundary perturbation theory~\cite{henry2005boundary} to derive the asymptotic approximation of eq.~\eqref{imageSphereAsymptotic}. 

\begin{figure}
    \centering
    \includegraphics[width=\linewidth]{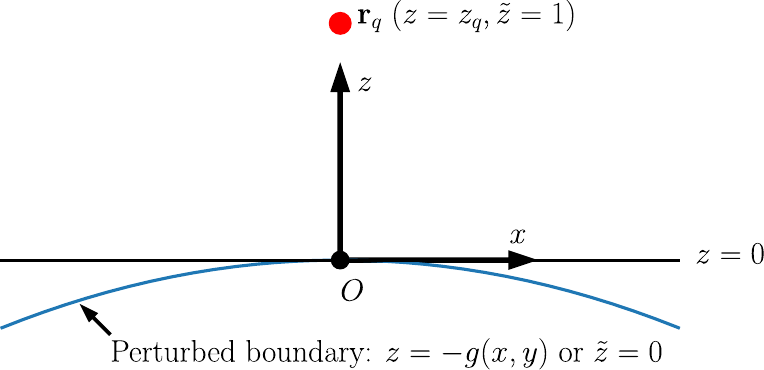}
    \caption{Schematic of the considered geometry and coordinate system (projection to the $x-z$ plane).}
    \label{schematicImage}
\end{figure}

To apply the standard boundary perturbation method \cite{henry2005boundary}, a new curvilinear coordinate system, for which the boundary condition is that of the unperturbed system needs to be defined. 
Furthermore, it is more convenient to work in a coordinate system that has been rescaled with respect to $z_q$.
Thus, I define the following curvilinear coordinate system, which is also depicted in Fig.~\ref{schematicImage}
\begin{equation} 
    \label{eq:variableTransform}
    \Tilde{x} = \frac{x}{z_q} \textrm{, } \Tilde{y} = \frac{y}{z_q} \textrm{, } \Tilde{z} = \frac{z + g(x,y)}{z_q} \textrm{.}
\end{equation}
The boundary condition in this frame simplifies into $\Phi(\Tilde{x},\Tilde{y},\Tilde{z} = 0)=0$; yet, the Poisson equation becomes significantly more complex than~\eqref{eq:Poisson}.

To write the Poisson equation in the $\Tilde{x},\Tilde{y}, \Tilde{z}$ coordinates, I apply the chain rule, along with the properties of the $\delta$ function, yielding
\begin{equation}
    \label{eq:PoissonCurvilinear}
    \begin{aligned}
          & \Phi_{\Tilde{x}\Tilde{x}} + \Phi_{\Tilde{y}\Tilde{y}} +  \Phi_{\Tilde{z}\Tilde{z}} \left(1 + g_{x}^2 + g_y^2\right) + \\
           & z_q \Phi_{\Tilde{z}} \left(g_{xx} + g_{yy} \right)  + 2 g_{x} \Phi_{\Tilde{x}\Tilde{z}} + 2 g_{y} \Phi_{\Tilde{y}\Tilde{z}} = \\
           &- \frac{q}{ z_q \epsilon_0} \delta(\Tilde{x}) \delta(\Tilde{y}) \delta(\Tilde{z}-1) \textrm{.}
    \end{aligned}
\end{equation}
Now I take the Taylor expansion of $g(x,y)$ around O and express it as a function of $\Tilde{x}, \Tilde{y}$; using the fact that the $x,y$ axes have been chosen to align with the principal axes of the surface, it yields 
\begin{equation}
    \label{boundaryRescaled}
    g(\Tilde{x},\Tilde{y}) = \frac{z_q}{2} \left(g_{xx} \Tilde{x}^2 + g_{yy} \Tilde{y}^2 \right)  + \mathcal{O} (z_q)^2  \textrm{, } z_q \rightarrow 0 \textrm{.}
\end{equation}
By introducing the small perturbation variable $\lambda = z_q g_{xx} / 2$, the above equation becomes 
\begin{equation}
    \label{boundaryRescaled}
    g(\Tilde{x}, \Tilde{y}) = \lambda \left(\Tilde{x}^2 + \kappa \Tilde{y}^2 \right) + \mathcal{O} (\lambda)^2  \textrm{, } \lambda \rightarrow 0 \textrm{,}
\end{equation}
where $\kappa = g_{yy} / g_{xx} < 1$ (assuming without loss of generality that the $x$ principal axis is the one with the larger curvature) is the ratio between the principal curvatures of the surface.
The small dimensionless perturbation parameter $\lambda$ is a metric of the proximity of the charge $q$ to the surface, in relation to its maximum local curvature. 

Substituting eq.~\eqref{boundaryRescaled} into \eqref{eq:PoissonCurvilinear} yields
\begin{equation}
    \label{eq:PoissonCurvilinear}
    \begin{aligned}
          & \Tilde{\nabla}^2 \Phi + \lambda \left( 2(1 + \kappa)\Phi_{\Tilde{z}}  + 4 \Tilde{x} \Phi_{\Tilde{x}\Tilde{z}} + 4 \kappa \Tilde{y} \Phi_{\Tilde{y}\Tilde{z}}  \right) + \\ & \mathcal{O}(\lambda)^2 =         - \frac{q}{ z_q \epsilon_0} \delta(\Tilde{x}) \delta(\Tilde{y}) \delta(\Tilde{z}-1) \textrm{,}
    \end{aligned}
\end{equation}
where $\Tilde{\nabla}$ denotes derivatives with respect $\Tilde{x}, \Tilde{y}, \Tilde{z}$.
I will now follow the standard perturbation method of expanding $\Phi$ in an asymptotic power series of $\lambda$
\begin{equation}
    \label{eq:asymptotic}
    \Phi = \Phi_0 + \lambda \Phi_1 + \mathcal{O}(\lambda)^2
\end{equation}
and match the terms of the same order.

Matching the zero-order terms yields
\begin{equation} \label{eq:zeroOrderEquation}
              \Tilde{\nabla}^2 \Phi_0  = - \frac{q}{ z_q \epsilon_0} \delta(\Tilde{x}) \delta(\Tilde{y}) \delta(\Tilde{z}-1) \textrm{,}
\end{equation}
with its solution being that of the well-known planar image point charge, i.e.,
\begin{equation} \label{eq:Phi0}
\begin{aligned}
        \Phi_0 = & \frac{q}{4 \pi  \epsilon_0 z_q} \left( \frac{1}{\sqrt{\Tilde{x}^2 + \Tilde{y}^2 + (\Tilde{z}-1)^2}} - \right.  \\
        & \left. \frac{1}{\sqrt{\Tilde{x}^2 + \Tilde{y}^2 + (\Tilde{z}+1)^2}} \right) \textrm{,} 
\end{aligned}
\end{equation}
as expected.
Matching the first-order terms in a similar fashion gives
\begin{equation} \label{eq:Phi1}
\begin{aligned}
    & - \Tilde{\nabla}^2 \Phi_1  =  2(1 + \kappa) \frac{\partial \Phi_0}{\partial \Tilde{z}}  + \\
   & + 4 \Tilde{x} \frac{\partial^2 \Phi_0}{\partial \Tilde{x} \partial \Tilde{z}} + 4 \kappa \Tilde{y} \frac{\partial^2 \Phi_0}{\partial \Tilde{y} \partial \Tilde{z}} = \frac{q}{2 \pi  \epsilon_0 z_q} f(\Tilde{x}, \Tilde{y}, \Tilde{z}) \textrm{,}
\end{aligned}
\end{equation}
where $f(\Tilde{x}, \Tilde{y}, \Tilde{z})$ can be calculated directly by substituting~\eqref{eq:Phi0} into~\eqref{eq:Phi1}
\begin{equation}
    \label{eq:Phi1RHS}
    \begin{aligned}
         f(\Tilde{x}, \Tilde{y}, \Tilde{z}) = & 
           6 \left(\Tilde{x}^2 + \kappa  \Tilde{y}^2 \right) \left[ \frac{\Tilde{z} - 1}{\left( \Tilde{x}^2 +  \Tilde{y}^2 + (\Tilde{z}-1)^2 \right)^{5/2}} \right. \\
          & \left. - \frac{\Tilde{z} + 1}{\left( \Tilde{x}^2 +  \Tilde{y}^2 + (\Tilde{z}+1)^2 \right)^{5/2}} \right] \\
          & + (1 + \kappa) \left[ \frac{\Tilde{z} + 1}{\left( \Tilde{x}^2 +  \Tilde{y}^2 + (\Tilde{z}+1)^2 \right)^{3/2}} \right. \\
          & \left.  - \frac{\Tilde{z} - 1}{\left( \Tilde{x}^2 +  \Tilde{y}^2 + (\Tilde{z}-1)^2 \right)^{3/2}} \right] \textrm{.}
    \end{aligned}
\end{equation}
To solve eq.~\eqref{eq:Phi1}, the Green's function for the Laplace operator can be utilized, while the boundary condition $\Phi_1(\Tilde{x}, \Tilde{y},\Tilde{z} = 0) = 0$ can be enforced by adding the contribution of the image reflection of $f(\Tilde{x}, \Tilde{y}, \Tilde{z})$ with respect to the $\Tilde{z}=0$ plane.

To obtain the potential energy of the image interaction from eq.~\eqref{eq:ImageInteraction}, I need to calculate $\Phi_1(\mathbf{r}_q) = \Phi_1(\Tilde{x}=0,\Tilde{y}=0,\Tilde{z}=1)$, which can be found by integrating $f$ and its image with respect to the $\Tilde{z}=0$ plane as
\begin{equation} \label{eq:Phi1SolutionSimple}
\begin{aligned}
        \Phi_1 (0,0,1)= & \frac{q}{8 \pi^2  \epsilon_0 z_q} 
     \left( \int_0^\infty h(\Tilde{z}) d\Tilde{z} - \int_{-\infty}^0 h(-\Tilde{z}) d\Tilde{z} \right)  \\
         = \frac{q}{8 \pi^2  \epsilon_0 z_q} & \left( \int_0^\infty h(\Tilde{z}) d\Tilde{z} + \int_0^{-\infty} h(\Tilde{z}) d\Tilde{z} \right)\textrm{,}
\end{aligned} 
\end{equation}
where
\begin{equation} \label{eq:XYintegral}
    h(\Tilde{z})=\int_{-\infty}^\infty \int_{-\infty}^\infty \frac{f(\Tilde{x}, \Tilde{y}, \Tilde{z})}{\sqrt{\Tilde{x}^2 + \Tilde{y}^2 + (\Tilde{z}-1)^2}}d\Tilde{x} d\Tilde{y}
\end{equation}
and in~\eqref{eq:Phi1SolutionSimple} I have simplified by considering that $f$ is even with respect to $\Tilde{z}$, i.e., $f(\Tilde{x}, \Tilde{y}, \Tilde{z}) = f(\Tilde{x}, \Tilde{y},- \Tilde{z})$. 
Integrals \eqref{eq:Phi1SolutionSimple} and \eqref{eq:XYintegral} are calculable analytically through laborious calculations performed using Wolfram Mathematica.
The corresponding notebook file is available on-line in~\cite{mathematicaNotebook} and is also attached as a PDF export in the supplementary material. 
It yields 
\begin{equation} \label{eq:Phi1Final}
    \Phi_1(0,0,1) = \frac{q}{16 \pi \epsilon_0 z_q}(1 + \kappa) \textrm{.}
\end{equation}

Combining eqs.~\eqref{eq:asymptotic} and \eqref{eq:Phi0} gives
\begin{equation}
    \label{eq:PhiAtPoint}
    \begin{aligned}
    \Phi_i(\mathbf{r}_q)&  = \frac{-q}{8 \pi \epsilon z_q} + \frac{q \lambda (1+ \kappa)}{16 \pi \epsilon_0 z_q} + \mathcal{O}(\lambda)^2 =  \\
    & \frac{-q}{8 \pi \epsilon_0 z_q} \left( 1 - \frac{z_q}{2R_m} +\mathcal{O}(\lambda)^2 \right)\textrm{.}
    \end{aligned} 
\end{equation}
Substituting back to eq.~\eqref{eq:ImageInteraction} and replacing the arbitrary point charge with the electron charge, i.e. $q = -e$, I obtain the central result of this section
\begin{equation}
    U_i(z_q) = -\frac{Q}{z_q} \left( 1 - \frac{z_q}{2R_m} +\mathcal{O}\left(\frac{z_q}{R_m}\right)^2 \right) \textrm{, } z_q \ll R_m \textrm{.}
\end{equation}

This result, along with the one from eq.~\eqref{eq:central1} form the main outcome of this paper, i.e. the proof that the eq.~\eqref{current} is valid for any surface shape, with the appropriate value for $R$ being the radius of mean curvature of the surface $R_m$.

\section{Conclusions}
In conclusion, this paper generalizes the results of Ref.~\cite{KXnonfn}, showing that the asymptotic approximations used in it are valid for emitting surfaces of any shape. The curvature-related correction terms of the expansion of the potential barrier along a path perpendicular to an arbitrary equipotential surface are proportional to the local mean curvature of the surface, i.e., the average of its two principal curvatures. This general result can be used to calculate, with mathematical accuracy, electron emission from surfaces of any geometry without having to extract the entire potential distribution in the tunneling region. Finally, it corrects a misconception in the literature~\cite{biswas_tunneling_2018} that connects the quadratic term to the second principal curvature of the surface. 

\section*{Supplementary Material}
The supplementary material contains an export of the Mathematica Notebook (available online in \cite{mathematicaNotebook}) that gives the details of calculating the integral of eq. \eqref{eq:Phi1SolutionSimple}, \eqref{eq:XYintegral}.

\section*{Acknowledgment}
This work was funded by the European Union’s Horizon 2020 research and innovation program, under grant agreement No 856705 (ERA Chair "MATTER").

\bibliographystyle{apsrev4-1}
\bibliography{bibliography/bibliography}

 \end{document}